\documentclass[12pt,times]{article}
\title{A dissipative particle dynamics method for arbitrarily complex geometries}

\author{Zhen Li\footnote{Email: \href{mailto:zhen_li@brown.edu}{zhen\_li@brown.edu}}, Xin Bian, Yu-Hang Tang and George Em Karniadakis\footnote{Email: \href{mailto:george_karniadakis@brown.edu}{george\_karniadakis@brown.edu}} \\
\small{Division of Applied Mathematics, Brown University, Providence, Rhode Island, 02912, USA} }

\date{\today}

\RequirePackage{lineno}
\usepackage[colorlinks=true]{hyperref}
\usepackage{graphicx}
\usepackage{multirow,array,color,mathrsfs,subcaption}
\usepackage{epstopdf} 

\usepackage{caption}
\DeclareGraphicsExtensions{.eps,.mps,.pdf,.jpg,.png}
\graphicspath{{figures/}{../figures/}}

\usepackage{fancyhdr}

\usepackage{amsmath}
\usepackage{amsfonts}

\usepackage{dsfont}

\usepackage[top=1.5cm,bottom=2.5cm,left=1.5cm,right=1.5cm]{geometry}

\usepackage[figuresright]{rotating}
\usepackage{extarrows}

\usepackage{fixltx2e}
\usepackage{xspace}
\newcommand{\usermeso}{\texttt{\textsubscript{\textit{USER}}MESO}\xspace}

\usepackage{amsmath,amssymb}
\begin{document}

\maketitle

\begin{abstract}
Dissipative particle dynamics (DPD) is an effective Lagrangian method for modeling complex fluids in the mesoscale regime but so far it has been limited to relatively simple geometries. Here, we formulate a local detection method for DPD involving arbitrarily shaped geometric three-dimensional domains. By introducing an indicator variable of boundary volume fraction (BVF) for each fluid particle, the boundary of arbitrary-shape objects is detected on-the-fly for the moving fluid particles using only the local particle configuration. Therefore, this approach eliminates the need of an analytical description of the boundary and geometry of objects in DPD simulations and makes it possible to load the geometry of a system directly from experimental images or computer-aided designs/drawings. More specifically, the BVF of a fluid particle is defined by the weighted summation over its neighboring particles within a cutoff distance. Wall penetration is inferred from the value of the BVF and prevented by a predictor-corrector algorithm. The no-slip boundary condition is achieved by employing effective dissipative coefficients for liquid-solid interactions. Quantitative evaluations of the new method are performed for the plane Poiseuille flow, the plane Couette flow and the Wannier flow in a cylindrical domain and compared with their corresponding analytical solutions and (high-order) spectral element solution of the Navier-Stokes equations. We verify that the proposed method yields correct no-slip boundary conditions for velocity and generates negligible fluctuations of density and temperature in the vicinity of the wall surface. Moreover, we construct a very complex 3D geometry -- the ``Brown Pacman" microfluidic device -- to explicitly demonstrate how to construct a DPD system with complex geometry directly from loading a graphical image. Subsequently, we simulate the flow of a surfactant solution through this complex microfluidic device using the new method. Its effectiveness is demonstrated by examining the rich dynamics of surfactant micelles, which are flowing around multiple small cylinders and stenotic regions in the microfluidic device without wall penetration. In addition to stationary arbitrary-shape objects, the new method is particularly useful for problems involving moving and deformable boundaries, because it only uses local information of neighboring particles and satisfies the desired boundary conditions on-the-fly.
\end{abstract}

\section{Introduction}
Despite of the sustained fast growth of computing power during the past few decades, it is still computationally prohibitive or impractical to model long time scales and large spatial scales in many applications of soft matter and biological systems with the brute-force atomistic simulations~\cite{2013Noid,2013Brini}. If only the mesoscopic properties and collective behavior are of practical interest, it may not be necessary to explicitly take into account all the details of materials at the atomic/molecular level~\cite{2013Yip}. To this end, a coarse-graining approach eliminates fast degrees of freedom and drastically simplifies the dynamics on atomistic scales, while providing a cost-effective simulation path to capturing the correct properties of complex fluids at larger spatial and temporal scales beyond the capacity of conventional atomistic simulations~\cite{2013Mills}. In recent years, with increasing attention on the research of soft matter and biophysics~\cite{2013Voth}, coarse-grained (CG) modeling has become a rapidly expanding methodology especially in the simulations of polymers~\cite{2015ZLi_CC,2016Lisal,2016D_Pan}, colloidal suspensions~\cite{2012Bian,2014Winkler,2014Bolintineanu}, interfaces of multiphase fluids~\cite{2013ZLi,2015Wang,2016Yang}, cell dynamics~\cite{2014Ye,2016Pivkin,2016Chang}, blood rheology~\cite{2013Lei,2016Henry,2016XJLi} and biological materials~\cite{2012XJLi,2014Zavadlav,2016Tang}.

Initially proposed by Hoogerbrugge and Koelman~\cite{1992Hoogerbrugge}, dissipative particle dynamics (DPD) is one of the currently most popular CG methods~\cite{2014Fedosov,2015Liu} for performing mesoscopic simulations of complex fluids. The DPD particles are defined as coarse-grained entities~\cite{2014ZLi_SM,2015ZLi_JCP}, which represent clusters of molecules rather than atoms/molecules directly. In contrast to molecular dynamics (MD) method, DPD allows much larger particle size and also time steps because of the soft particle interactions. As a particle-based mesoscopic method, DPD considers $N$ particles, whose state variables of momentum and position are governed by the Newton's equations of motion~\cite{1997Groot}. For a typical DPD particle $i$, its time evolution follows $\mathbf{\dot{r}}_i=\mathbf{v}_{i}$ and ${\mathbf{\dot{p}}_i}=\mathbf{F}_{i}=\sum_{i\neq j}(\mathbf{F}_{ij}^{C} + \mathbf{F}_{ij}^{D} + \mathbf{F}_{ij}^{R})$ where $\mathbf{r}_i$, $\mathbf{v}_i$, $\mathbf{p}_i$ and $\mathbf{F}_i$ denote position, velocity, momentum and force vectors, respectively. The summation for computing the total force $\mathbf{F}_i$ is carried out over all other particles within a cutoff radius $r_c$ beyond which the forces are considered negligible. The pairwise force $\mathbf{F}_{ij}$ comprises conservative ($\mathbf{F}_{ij}^{C}$), dissipative ($\mathbf{F}_{ij}^{D}$) and random ($\mathbf{F}_{ij}^{R}$) forces are expressed as~\cite{1997Groot}
\begin{equation}\label{equ:DPD_force}
\begin{split}
  & \mathbf{F}_{ij}^{C} =  a_{ij}{\omega_{C}}(r_{ij})\mathbf{e}_{ij} ,\\
  & \mathbf{F}_{ij}^{D} = -\gamma_{ij} {\omega_{D}}(r_{ij})(\mathbf{e}_{ij} \cdot \mathbf{v}_{ij})\mathbf{e}_{ij} ,\\
  & \mathbf{F}_{ij}^{R} = \sigma_{ij} {\omega_{R}}(r_{ij}) {\rm d}{\tilde W}_{ij}\mathbf{e}_{ij} ,
\end{split}
\end{equation}
where $r_{ij}=|\mathbf{r}_{ij}|=|\mathbf{r}_{i}-\mathbf{r}_j|$ represents the distance between two particles $i$ and $j$, $\mathbf{e}_{ij}=\mathbf{r}_{ij}/r_{ij}$ is the unit vector from particles $j$ to $i$, and $\mathbf{v}_{ij}=\mathbf{v}_i-\mathbf{v}_j$ is the velocity difference; ${\rm d}{\tilde W}_{ij}$ is an independent increment of the Wiener process~\cite{1995Espanol}. Also, $\gamma_{ij}$ is the dissipative coefficient and $\sigma_{ij}$ sets the strength of random force. The dissipative force and random force together act as a thermostat when the dissipative coefficient $\gamma$ and the amplitudes of white noise $\sigma$ satisfy the fluctuation-dissipation theorem (FDT)~\cite{1966Kubo,1995Espanol} requiring $\sigma^2=2\gamma k_B T$ and $\omega_D(r)=\omega_R^2(r)$. All these forces in Eq.~\eqref{equ:DPD_force} have the same finite interaction range $r_c$ and their amplitudes decay according to corresponding weight functions. A common choice of the weight functions~\cite{1997Groot} is $\omega_C(r)=1-r/r_c$ and $\omega_D(r)=\omega_R^2(r)=\left(1-r/r_c\right)^2$ for $r \leq r_c$ and zero for $r>r_c$.

All the three forces between DPD particles are soft and short-range interactions, which allows large time steps for the time integration of the particle-based system. The soft interactions between DPD particles, unlike the hard potentials in atomistic simulations, cannot prevent fluid particles from penetrating wall boundaries~\cite{1989Allen}. It is also unlike the top-down smoothed particle hydrodynamics (SPH)~\cite{2012Adami} or smoothed DPD (SDPD)~\cite{2012Bian} approach, where the equation of state can be tuned so that the pressure is arbitrarily strong to prevent particle penetration. As a result, for wall-bounded flow systems, DPD simulations require extra formulations~\cite{2005Pivkin,2006Colmenares,2007Altenhoff} to prevent the penetration of the liquid particles into solid boundaries. Specular, Maxwellian, and bounce-back reflections~\cite{1999Revenga} are common techniques used to reflect particles back into the fluid after they cross the wall surface. Therefore, for wall-bounded flows one has to mathematically predefine the position of solid wall to judge the penetration of fluid particles before a DPD simulation can be performed, which is difficult to extend for arbitrarily shaped boundaries and limits the applicability of DPD.

In the present paper, we develop a boundary method for imposing correctly the no-slip boundary condition on the solid walls with arbitrary shapes. Instead of predefining the position of the wall boundary, we make the fluid particles autonomous to detect the wall surface and to infer the wall penetration by themselves based on the local information of their neighboring particles. Hence, the geometry of solid boundary can be computed on-the-fly using local particle configurations. Therefore, it is no longer necessary to predefine the boundary geometry for DPD simulations, which makes it possible to construct DPD systems with arbitrary-shape domains directly from loading experimental images or computer-aided designs/drawings. Furthermore, since this boundary method uses local information of neighboring particles and satisfies no-slip/partial-slip boundary conditions on-the-fly, it is not only valuable for stationary arbitrary-shape boundaries but also for moving boundaries and deformable boundaries.

The remainder of this paper is organized as follows: Section~\ref{sec:2} introduces the details of the boundary method, and also how to compute the effective dissipative coefficient for liquid-solid interactions. In Section~\ref{sec:3}, we validate the proposed boundary method by performing the Poiseuille flow, the Couette flow and the Wannier flow with comparison to analytical solutions. Moreover, an error analysis of this boundary method related to the curvature of arbitrary-shaped boundaries is provided in Appendix~\ref{sec:app}. We also include a demonstration of micelles flowing through  a very complex microfludic device. Finally, we end with a brief summary and discussion in Section~\ref{sec:4}.

\section{Wall Boundary Method}\label{sec:2}
\subsection{Definition of the boundary volume fraction}
Consider a fluid particle $i$ in the vicinity of a solid wall represented by discrete DPD particles and we assign to it an extra variable $\phi_i=\varphi(\mathbf{r}_{i})$ in addition to other quantities such as position and momentum. We define $\phi_i$ as the boundary volume fraction (BVF) depending on the coordinates of particle $i$. More specifically, the value of $\phi_i$ is computed using a weighted summation over neighboring solid particles $j$ given by
\begin{equation}\label{eq:Phi}
  \phi_i = \varphi(\mathbf{r}_{i})=\frac{1}{\rho_w}\sum_{r_{ij} < r_{cw}}^{j\in \mathbf{S}}W(r_{ij},r_{cw}) ~,
\end{equation}
\noindent
where $W(r,r_{cw})$ is a weighting function, and $\rho_w$ is the bulk number density of solid particles. The weighting function $W(r,r_{cw})$ can be any smoothing kernel, such as the ones used widely in smoothed particle hydrodynamics~\cite{2003Liu,2005Monaghan}. As a demonstration, we choose the three-dimensional Lucy kernel~\cite{1977Lucy}
\begin{equation}\label{eq:Lucy}
  W(r,r_{cw})= \frac{105}{16\pi r^3_{cw}}\left(1+\frac{3r}{r_{cw}}\right)\left(1-\frac{r}{r_{cw}}\right)^3 ~,
\end{equation}
\noindent
where $r$ is the norm of $\mathbf{r}$, and $r_{cw}$ is the cutoff radius beyond which $W(r,r_{cw})$ is considered zero. Larger $r_{cw}$ increases the computational cost but yields smoother $\varphi(\mathbf{r})$, as we will discuss in section \ref{sec:3}. Unless otherwise specified, in testing cases we simply set $r_{cw}$ equal to $r_c$.

\begin{figure}[b!]
  \centering
  \includegraphics[width=0.8\textwidth]{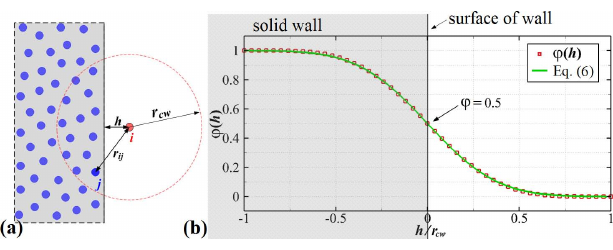}
  \caption{(a) Sketch for a fluid particle $i$ in the vicinity of a wall represented by many solid particles and the integration domain for accumulation of $\phi$. (b) Dependence of $\phi(h)$ on $h/r_{cw}$ calculated by the integration in Eq. (\ref{eq:Estimate}).}
  \label{fig:Geo}       
\end{figure}

Consider a planar wall surface or a wall surface with a radius of curvature far greater than the cutoff radius $r_{cw}$, as shown in Fig.~\ref{fig:Geo}(a); we estimate the value of $\phi_i$ using the continuum approximation
\begin{equation}\label{eq:Estimate}
  \phi_i=\varphi(h)= \int_{z = h}^{r_{cw}} \int_{x = 0}^{\sqrt{{r_{cw}}^2 - z^2}} 2\pi x W(r,r_{cw}) \cdot dx \cdot dz
\end{equation}
\noindent
where $r=\sqrt{x^2+z^2}$ and $h$ is the distance between the particle $i$ and the wall boundary. By inserting the Lucy kernel given by Eq. (\ref{eq:Lucy}) into Eq. (\ref{eq:Estimate}), we have

\begin{equation}\label{eq:phi_h}
  \varphi(h)=
  \left\{\begin{matrix}
        &\frac{1}{16}\left(1-\frac{h}{r_{cw}}\right)^5 \left( 15\left(\frac{h}{r_{cw}}\right)^2+19\frac{h}{r_{cw}}+8  \right)  &0\leqslant h\leqslant r_{cw}\ , \\
        &1-\varphi(-h)  &-r_{cw}\leqslant h<0
  \end{matrix}\right.
\end{equation}
\noindent
in which $\varphi(h=0)=0.5$. It is worth noting that $h=0$ represents that the particle lies right on the wall surface, and a negative $h$ means that the particle is underneath the wall surface while a positive $h$ is for the particle outside the wall boundary. The number density $\rho_w$ disappears in Eq.~\eqref{eq:Estimate} because we scaled $\varphi(\mathbf{r})$ by $\rho_w$ in Eq.~\eqref{eq:Phi}. Thus, the value of $\phi_i$ only depends on $h$ as plotted in Fig.~\ref{fig:Geo}(b), which shows clearly that $\varphi(h)$ decreases from $1.0$ to $0$ as $h$ changes from $-r_{cw}$ to $r_{cw}$.

Given a value of $\phi_i$, the distance of a particle $i$ away from the wall surface can be computed by the inverse function of Eq.~\eqref{eq:phi_h}. For simplifying the numerical implementation in practical simulations, we employ an approximation of $\varphi^{-1}(\phi_i)$ in the form of

\begin{equation}\label{eq:inverse_eq}
  h_i/r_{cw} = \varphi^{-1}(\phi_i) \approx
  \left\{\begin{matrix}
        &1-\left(2.088\phi_i^3 + 1.478\phi_i\right)^{1/4}  &0\leqslant \phi_i\leqslant 0.5\ , \\
        &-\varphi^{-1}(1-\phi_i)  &0.5< \phi_i\leqslant1 .
  \end{matrix}\right.
\end{equation}

The value of BVF on each fluid particle $\phi_i$ can be evaluated every time step in the same loop of pairwise force computation, so the extra computational cost for applying this boundary method is marginal. Given a value of $\phi_i$, the distance of the particle $i$ from the wall surface is given by $h=\varphi^{-1}(\phi_i)\cdot r_{cw}$. Whenever $h_i<0$ (or $\phi_i>0.5$), the wall penetration is observed.

For a fluid DPD particle $i$ with state variables $\{\mathbf{r}_i,\mathbf{v}_i\}$, we employ a predictor-corrector algorithm to prevent the wall penetration. In particular, we perform an imaginary-integration of its position for a time step $\Delta t$, i.e., $\mathbf{r}'_i=\mathbf{r}_i+\mathbf{v}_i\Delta t$. If the value $\varphi(\mathbf{r}'_i)>0.5$, the particle $i$ at the predicted position $\mathbf{r}'_i$ would penetrate into the wall. To avoid this wall penetration, we correct the velocity of the particles whose $\varphi(\mathbf{r}'_i)$ greater than $0.5$ by reassigning a new value
\begin{equation}\label{equ:bounce_back}
\mathbf{v}^{\rm new}_i=2\mathbf{U}+\mathbf{a}\Delta t-\mathbf{v}_i + 2~{\rm max}\{0,\mathbf{v}_i \cdot \mathbf{e}_n\}\cdot \mathbf{e}_n,
\end{equation}
where $\mathbf{e}_n$ is the unit normal vector of the wall boundary, $\mathbf{U}$ and $\mathbf{a}$ are the local velocity and local acceleration of the boundary, respectively. It is obvious that all stationary wall boundaries have $\mathbf{U}=\mathbf{a}=0$. For moving boundaries, the value of $\mathbf{U}$ and $\mathbf{a}$ can take the values of velocity and acceleration of the nearest wall particle in practical DPD simulations. Let $\mathbf{n}_w$ be the gradient of $\varphi(\mathbf{r}_i)$ at the location of particle $i$, which is computed by
\begin{equation}\label{equ:gradient_phi}
\mathbf{n}_w = \nabla\varphi(\mathbf{r}_i)=\frac{1}{\rho_w}\sum_{r_{ij} < r_{cw}}^{j\in \mathbf{S}}\frac{\mathbf{r}_{ij}}{r_{ij}}\frac{{\rm d}W(r_{ij},r_{cw})}{{\rm d}r_{ij}} .
\end{equation}
Then, the unit normal vector of the wall boundary $\mathbf{e}_n = \mathbf{n}_w/{n}_w$ where $n_w$ is the modulus of $\mathbf{n}_w$.

\subsection{Control of the surface roughness}
Consider a flat wall represented by solid particles in uniform lattices, the iso-surface of $\phi=0.5$ is smooth and flat, which can accurately represent the surface of the flat wall. However, the structure of solid particles associated with these lattices will induce unwanted fluctuations~\cite{2005Pivkin} of fluid density and temperature in the vicinity of wall boundary. In the present paper, we employ randomly distributed particles in the wall domain to represent the wall boundary, which is easier and more general for construction of fluid system with complex geometry and arbitrarily shaped boundaries.

\begin{figure}[t!]
  \centering
  \includegraphics[width=0.84\textwidth]{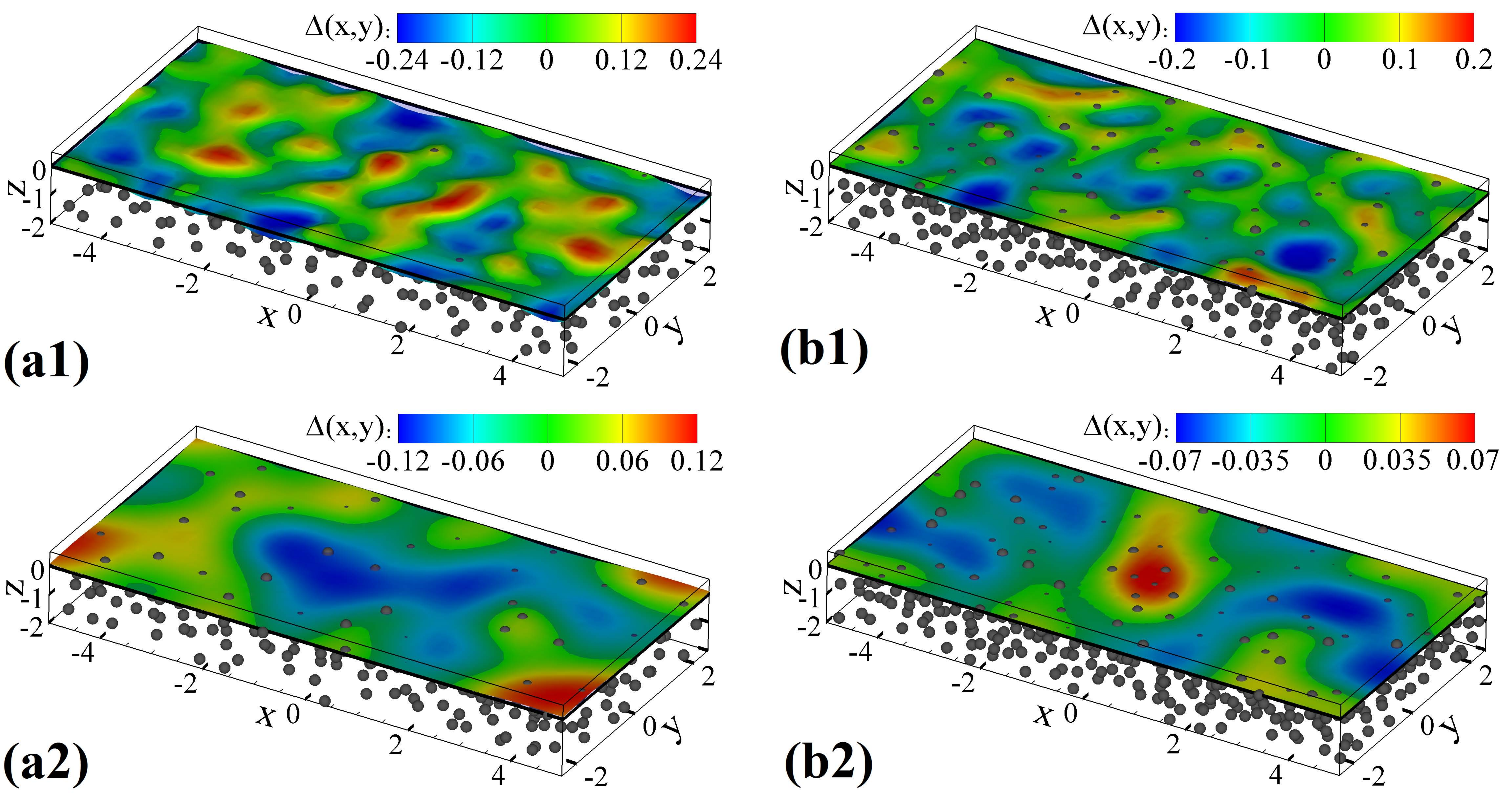}
  \caption{Wall boundary represented by the isosurface of $\phi=0.5$ computed by Eq.~\eqref{eq:Phi} for a number density of wall particles $\rho_w=4.0$ with the cutoff radius of (a1) $r_{cw}=1.0$ and (a2) $r_{cw}=2.0$, and for a number density $\rho_w=8.0$ with the cutoff radius of (b1) $r_{cw}=1.0$ and (b2) $r_{cw}=2.0$. The value of $\Delta(x,y)$ represents the vertical deviations of an isosurface of $\phi=0.5$ from its ideal surface $z=0$. Spherules represent the randomly distributed solid particles.}
  \label{fig:wall}       
\end{figure}

Randomly distributed particles do not possess a lattice structure, and hence effectively eliminate the fluctuations of averaged profiles of density and temperature in the vicinity of the wall surface. Theoretically, to the limit of the continuum approximation, a planar wall surface can be accurately represented by the isosurface of $\phi=0.5$ when the solid particles are dense enough. However, in practical implementations, the number density $\rho_w$ in DPD simulations is finite and usually smaller than $10.0$. As a result, the roughness on wall surface is generated by the random distribution of discrete particles with finite number density. Figures~\ref{fig:wall} shows the wall boundary represented by the isosurface of $\phi=0.5$ for the number densities of wall particles being $\rho_w=4.0$ and $8.0$. The solid wall is made up of randomly distributed particles, where the average distance between these particles is $\delta=\rho_w^{-1/3}=0.63$ for $\rho_w=4.0$ and $\delta=0.5$ for $\rho_w=8.0$.

The surface roughness shown in Fig.~\ref{fig:wall} comes from the estimation error of $\phi$ based on the discrete particles. Unlike SPH or other mesh-free methods that try to eliminate this error, we take advantage of this kind of error for generating controllable roughness on wall surface in the DPD systems. As a matter of fact, any natural solid wall contains more or less chemical/physical heterogeneities on the surface especially at the mesoscopic scale. Such heterogeneity can be modeled qualitatively by the roughness on walls in the DPD systems, as shown in Fig.~\ref{fig:wall}. More importantly, the proposed boundary method provides a convenient way to generate various sizes of the roughness for representation of different degrees of the heterogeneity.

\begin{figure}[h!]
  \centering
  \includegraphics[width=0.84\textwidth]{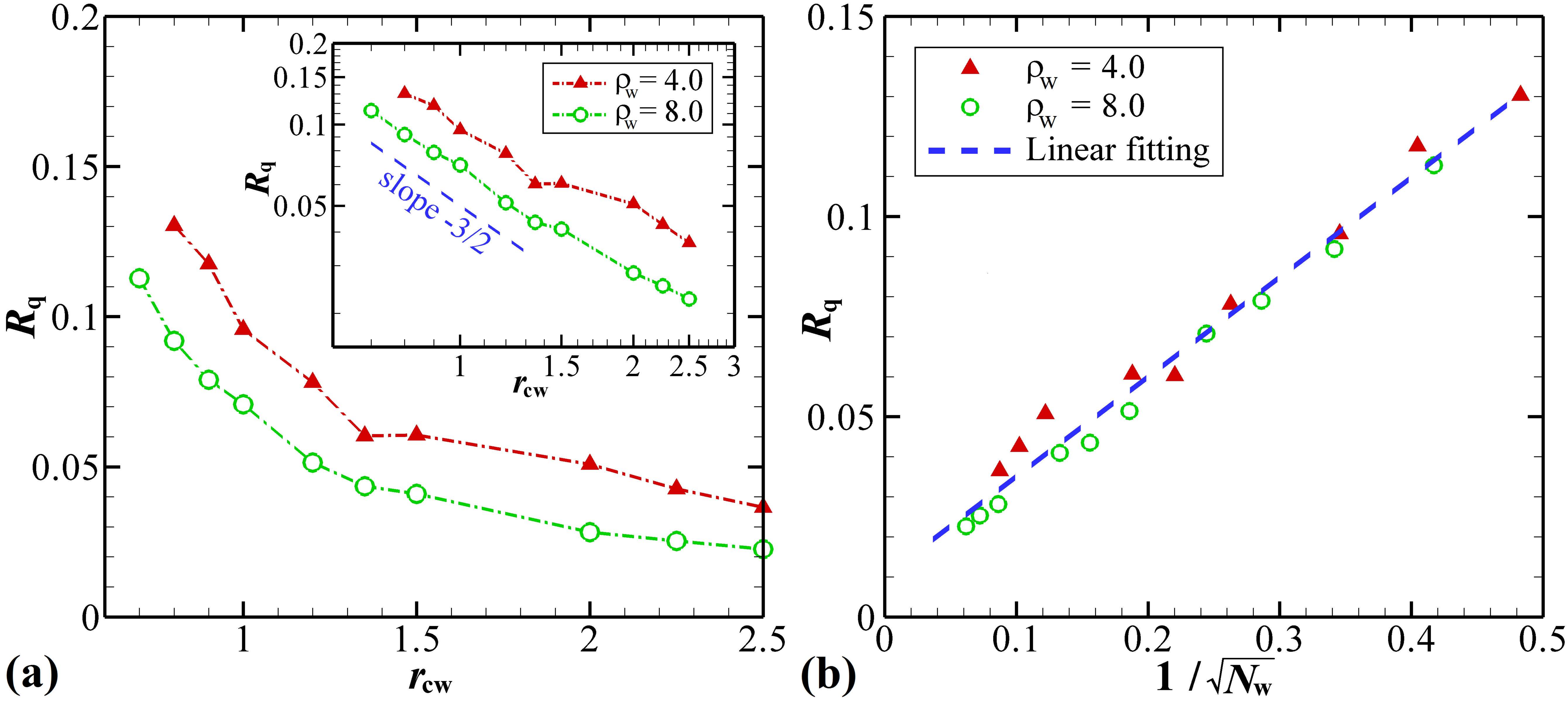}
  \caption{(a) Dependence of the root mean squared height $R_q$ on the cutoff radius $r_{cw}$, where the inset shows the scaling law of $R_q\sim r_{cw}^{-3/2}$. (b) The scaling law of $R_q\propto 1/\sqrt{N_w}$ indicating the linear tunability of the wall surface roughness, where $N_w$ is the number of neighboring solid particles on wall surface, i.e., $N_w=2\pi r_{cw}^3/3\cdot\rho_w$.}
  \label{fig:Rq}       
\end{figure}

According to the unbiased estimation of standard deviation~\cite{2011Kroese}, the magnitude of roughness associated with the randomness of particle distribution will monotonically decrease as the number of neighboring particles $N_w=2\pi r_{cw}^3/3\cdot\rho_w$ increases, implying that the roughness of the wall is controllable by tuning the cutoff radius $r_{cw}$ and number density $\rho_w$ of DPD systems. To quantify the wall surface texture, we define the root mean squared height $R_q$ given by
\begin{equation}\label{Eq:roughness}
  R_q=\left( \frac{1}{A}\iint \Delta^2(x,y) dxdy \right)^{1/2}
\end{equation}
\noindent
where $A$ is the area of a flat wall, and $\Delta(x,y)$ represents the vertical deviations of a real surface of $\phi=0.5$ from its ideal surface defined by the isosurface of $\phi=0.5$ when $\rho_w\to\infty$.
Figure~\ref{fig:Rq} shows that the root mean squared height $R_q$ decreases as the cutoff radius $r_{cw}$ and the number density $\rho_w$ increase. Since $r_{cw}$ is only involved in computation of $\phi$, the variation of $r_{cw}$ does not affect the particle interactions and fluid properties. In practical DPD simulations, changing the value of $N_w^{-1/2}$ allows linear tunability of the wall surface roughness, as shown in Fig.~\ref{fig:Rq}(b).

Usually, a randomized configuration of DPD particles can be generated simply by a random number generator, which may result in overlapping of particles or large vacancies in wall boundaries. To avoid the overlaps or vacancies, a more uniform particle distribution is needed, which can be achieved by a process of geometry optimization or a short run of particle-based simulation. In the present paper, we carry out a short DPD simulation with a relatively large conservative force coefficient to get the initial particle positions. Then, the particles in the wall domain are frozen as solid particles, while others in the fluid domain are taken as the fluid particles. For instance, the wall boundary shown in Fig.~\ref{fig:wall} is obtained by running $1000$ time steps of DPD simulation from totally randomized particles.

\subsection{Effective dissipative interaction}
\begin{figure}[b!]
  \centering
  \includegraphics[width=0.38\textwidth]{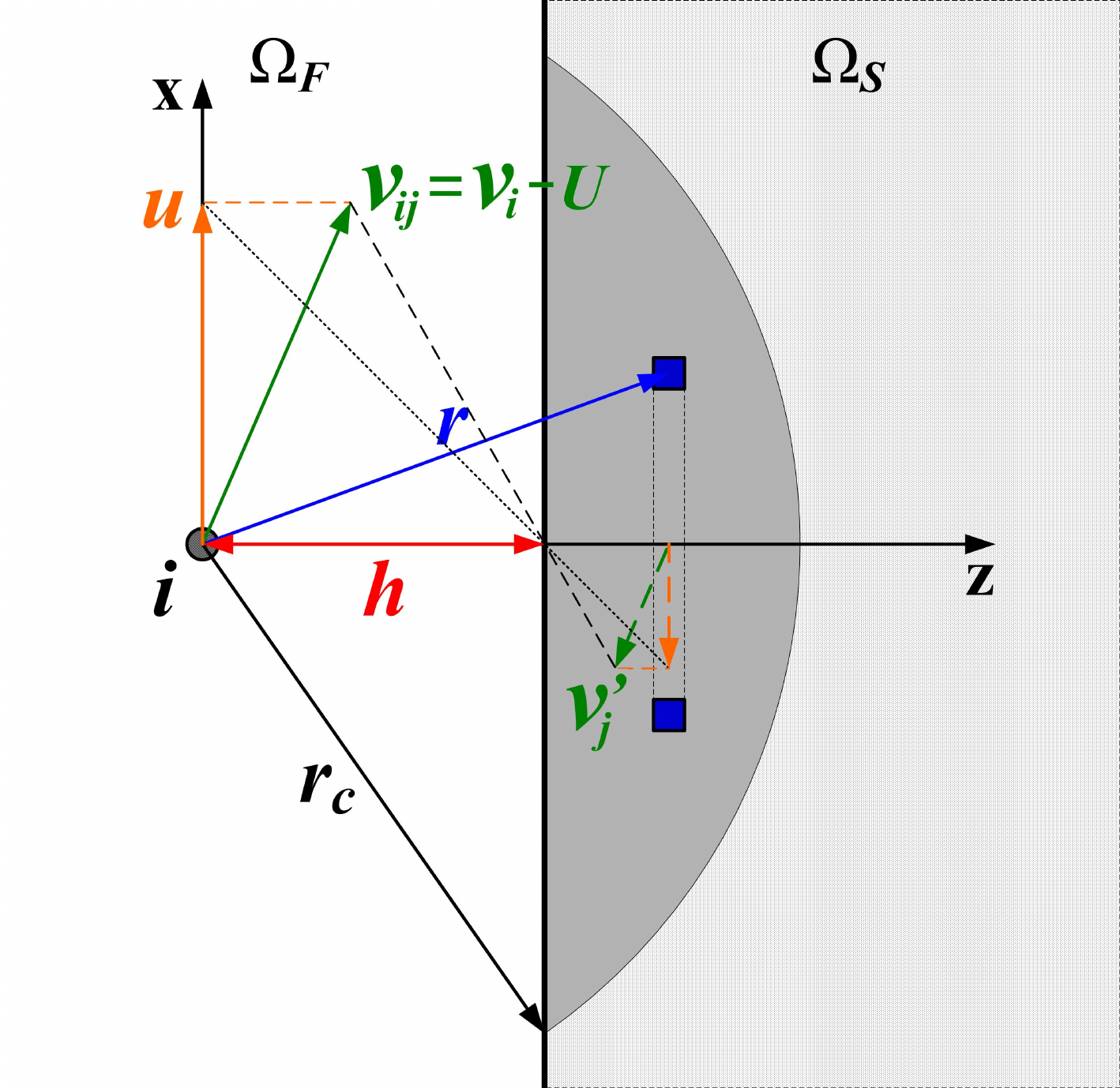}
  \caption{Integration domains for the effective boundary force and heat flux. $\Omega_F$ and $\Omega_S$ represent the domain of fluid and solid wall, respectively. The number of particles in the infinitesimal ring element is $2\pi \rho g(r)xdxdz$ in which $\rho$ is the average number density and $g(r)$ the radial distribution function. $\mathbf{v}_{ij}=\mathbf{v}_i-\mathbf{U}$ is the instantaneous velocity difference between particle $i$ and the boundary, while $\mathbf{v}'_{j}=-\mathbf{v}_{ij}\cdot (z-h)/h$ is an extrapolated velocity for solid particles.}
  \label{fig:FD}
\end{figure}

The dissipative force between two DPD particles is computed by $\mathbf{F}_{IJ}^{D} = -\gamma \cdot\omega_D(r_{ij})(\mathbf{e}_{ij} \cdot \mathbf{v}_{ij})\mathbf{e}_{ij}$, where $r_{ij}$ is the distance between particles $i$ and $j$, $\mathbf{e}_{ij}$ represents the unit vector from particle $j$ to $i$, and $\mathbf{v}_{ij}=\mathbf{v}_i-\mathbf{v}_j$ is their velocity difference. The weighting function is given by $\omega_D(r_{ij})=(1-r_{ij}/r_c)^2$ for $r_{ij}\leqslant r_c$ and zero for $r_{ij} > r_c$. The effective dissipative force from solid boundaries is extracted from the fluid-solid interactions using the continuum approximation~\cite{2014ZLi_JCP}. In particular, we integrate the force contribution over the part of the cutoff sphere that lies in the solid domain, as shown with the gray area of Fig.~\ref{fig:FD} where we choose a coordinate system with $x$-axis along the direction of the velocity component of particle $i$ parallel to the wall surface and $z$-axis perpendicular to the wall surface. Let $\mathbf{v}_{ij} = \mathbf{v}_i-\mathbf{U}$ be the instantaneous velocity difference between particle $i$ with velocity $\mathbf{v}_i$ and the solid particles with velocity $\mathbf{U}$, and $u$ be the parallel component of $\mathbf{v}_{ij}$.

Consider a planar wall surface or a wall surface with the radius of curvature far greater than the cutoff radius $r_c$; if the wall is considered as a rigid body and has uniform velocity $\mathbf{U}$, the instantaneous velocity difference in parallel direction is $u$. Then, the total dissipative force $F^D_0(h)$ on the particle $i$ due to the presence of wall boundary can be evaluated by:
\begin{eqnarray}\label{FD_int}
   F^D_0(h) &=& \int_{z = h}^{r_C} \int_{x = 0}^{\sqrt {{r_C}^2 - z^2}}\int_{\theta = 0}^{2\pi} \left(- \gamma \cdot\omega_D(r_{ij})(\mathbf{e}_{ij} \cdot \mathbf{v}_{ij})\mathbf{e}_{ij} \cdot \rho \cdot g(r) \cdot dx \cdot x \cdot d\theta \cdot dz \right) \nonumber\\
   &=& \int_{z = h}^{r_C} \int_{x = 0}^{\sqrt {{r_C}^2 - z^2}}\left( -\gamma\pi\rho \cdot u \cdot \left(1-\frac{r}{r_c}\right)^2\cdot\frac{x^3}{r^2}\cdot g(r)\cdot dx \cdot dz \right) \nonumber\\
   &\xlongequal{g(r)=1}& -\gamma\pi\rho u r_c^3\left[ \frac{1}{45}-\frac{1}{12}\frac{h}{r_c} -\left(\frac{h}{r_c}\right)^3\left(\frac{1}{3}\log\left(\frac{h}{r_c}\right)+\frac{2}{9}\right)+ \frac{1}{3}\left(\frac{h}{r_c}\right)^4 - \frac{1}{20}\left(\frac{h}{r_c}\right)^5 \right]
\end{eqnarray}

\begin{figure}[b!]
  \centering
  \includegraphics[width=0.9\textwidth]{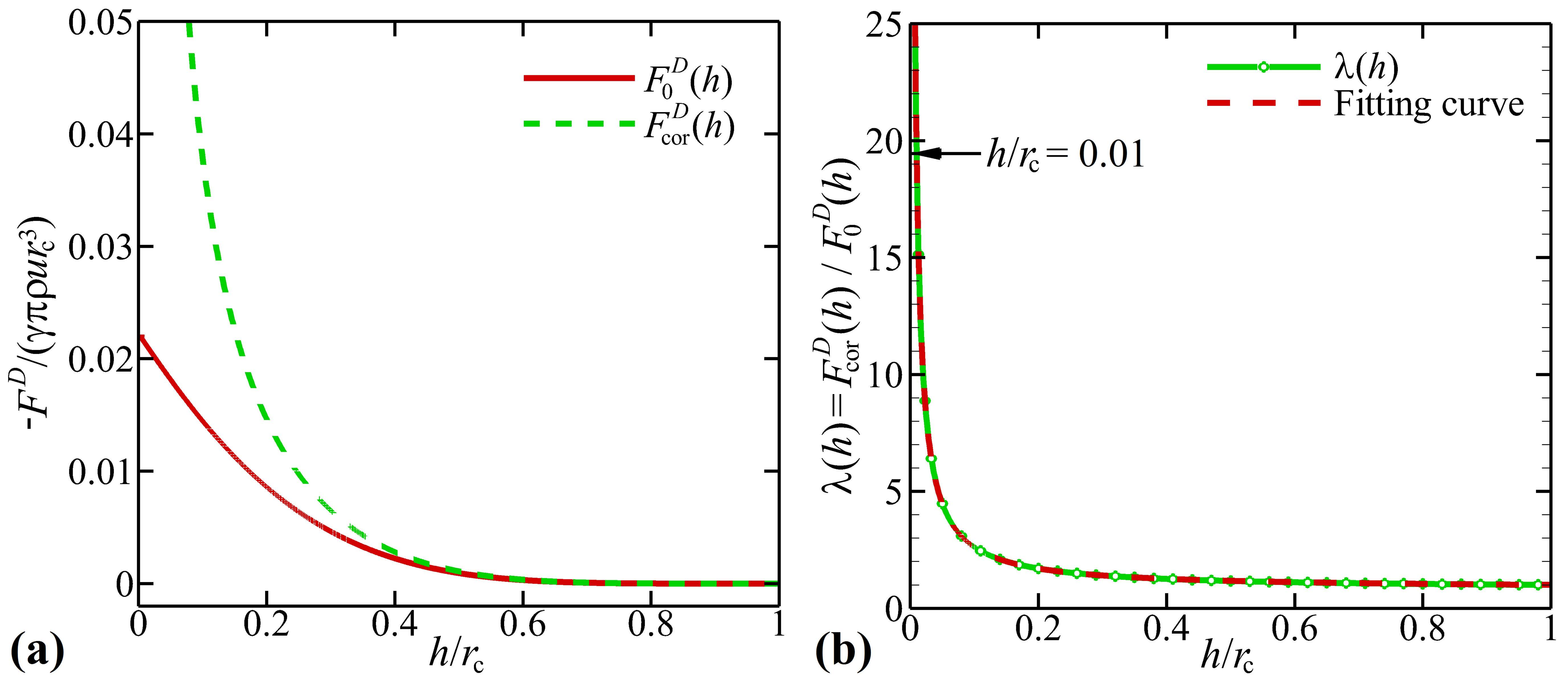}\\
  \caption{(a) Distance-dependent functions of the effective dissipative force $F^D_0(h)$ and $F^D_{\rm cor}(h)$. (b) Correction coefficient $\lambda(h)$ defined by the ratio of $F^D_{\rm cor}(h)$ to $F^D_0(h)$ and its approximation given by Eq.~\eqref{equ:lamda_fit}.}
  \label{fig:lamda_fit}
  \vspace{-5pt}
\end{figure}

However, the value of $F^D_0(h)$ is not sufficient to impose the correct no-slip boundary condition on the wall surface. To this end, we assign an extrapolated velocity $\mathbf{v}'_{j}=-\mathbf{v}_{ij}\cdot (z-h)/h$ to each solid particle so that the wall surface has zero velocity, and hence the instantaneous velocity difference becomes $\tilde{\mathbf{v}}_{ij}=\mathbf{v}_{ij}-\mathbf{v}'_{j}=\mathbf{v}_{ij}\cdot z/h$, in which the parallel component is $u\cdot z/h$. Then, the corrected dissipative force $F^D_{\rm cor}(h)$ on the particle $i$ due to the presence of wall boundary is computed by:
\begin{eqnarray}\label{FD_cor}
   F^D_{\rm cor}(h) &=& \int_{z = h}^{r_C} \int_{x = 0}^{\sqrt {{r_C}^2 - z^2}}\int_{\theta = 0}^{2\pi} \left(- \gamma \cdot\omega_D(r_{ij})(\mathbf{e}_{ij} \cdot \mathbf{\tilde{v}}_{ij})\mathbf{e}_{ij} \cdot \rho \cdot g(r) \cdot dx \cdot x \cdot d\theta \cdot dz \right) \nonumber\\
   &=& \int_{z = h}^{r_C} \int_{x = 0}^{\sqrt {{r_C}^2 - z^2}}\left( -\gamma\pi\rho \cdot \left(\frac{z}{h}u\right) \cdot \left(1-\frac{r}{r_c}\right)^2 \cdot \frac{x^3}{r^2}\cdot g(r)\cdot dx \cdot dz \right) \nonumber\\
   &\xlongequal{g(r)=1}& -\gamma\pi\rho u r_c^3\left[ \frac{1}{240}\frac{r_c}{h}-\frac{1}{24}\frac{h}{r_c}-\frac{1}{4}\left(\frac{h}{r_c}\right)^3 \left[ \log \left(\frac{h}{c} \right)+\frac{3}{4}\right]+\frac{4}{15}\left(\frac{h}{r_c}\right)^4-\frac{1}{24}\left(\frac{h}{r_c}\right)^5  \right]
\end{eqnarray}

Figure~\ref{fig:lamda_fit}(a) plots the distance-dependent functions of $F^D_0(h)$ and $F^D_{\rm cor}(h)$, which shows that the correction with extrapolated velocities does not change the value of $F^D(h)$ significantly at large distances (i.e., $h/r_c>0.5$) but yields bigger dissipative force at small distances (i.e., $h/r_c<0.5$). Here, we define the ratio of $F^D_{\rm cor}(h)$ to $F^D_0(h)$ as a correction coefficient $\lambda(h)=F^D_{\rm cor}(h)/F^D_0(h)$, which is plotted in Fig.~\ref{fig:lamda_fit}(b). In practical implementation, the distance-dependent coefficient $\lambda(h)$ can be approximated by
\begin{equation}\label{equ:lamda_fit}
  \lambda(h) =  \lambda(\varphi^{-1}(\phi)\cdot r_{cw}) \approx
  \left\{\begin{matrix}
        &1+0.187\left(\frac{r_c}{h}-1\right)-0.093\left(1-\frac{h}{r_c}\right)^3  &0.01 \leqslant h/r_c \leqslant 1.0\ , \\
        &19.423  & h/r_c < 0.01 \ .
  \end{matrix}\right.
\end{equation}
Let the effective dissipative coefficient for liquid-solid interaction be $\gamma_e = \lambda(h)\cdot \gamma$. We note that the formula of Eq.~\eqref{equ:lamda_fit} is obtained based on $g(r)=1$. A more accurate function of $\lambda(h)$ can be derived from Eqs.~\eqref{FD_int} and~\eqref{FD_cor} using the computed $g(r)$. For easier numerical implementation using Eq.~\eqref{equ:lamda_fit} directly without computation of $g(r)$, it is recommended to keep $N_w=2\pi r^3_{cw}/3\cdot \rho_w \geqslant 15$, i.e., setting $r_{cw}\geqslant 1.35$ at $\rho_w=3$ and $r_{cw}\geqslant1.0$ at $\rho_w=8$, so that the value of $\phi$ can be evaluated accurately. Then, the dissipative force between liquid particles and solid particles is computed by $\mathbf{F}_{IJ}^{D} = - \gamma_e\cdot\omega_D(r_{ij})(\mathbf{e}_{ij} \cdot \mathbf{v}_{ij})\mathbf{e}_{ij}$, which guarantees the no-slip boundary condition at the wall surface. The corresponding random force is given by $\mathbf{F}_{IJ}^{R} = \sigma_e\cdot \omega_R(r_{ij}) {\rm d}{\tilde W}{ij}\mathbf{e}_{ij}$ with $\sigma_e = 2k_BT\gamma_e$ and ${\rm d}{\tilde W}_{ij}$ being independent increments of the Wiener process to satisfy the FDT~\cite{1995Espanol}. In the next section, we will verify the validity and the accuracy of the boundary method using the effective dissipative coefficient for liquid-solid interaction.

\section{Numerical Results}\label{sec:3}
In this section, we examine the accuracy of the proposed boundary method for well-known flows such as the plane Poiseuille flow, the plane Couette flow and the Wannier cylindrical flow. Then, a demonstration of flow in a ``Brown Pacman" microfluidic device involving very complex boundaries is performed.

Firstly, we test the accuracy of the boundary method on stationary walls by carrying out a DPD simulation of the plane Poiseuille flow, in which a body force field acting in the $x$-direction on a fluid between two flat plates in the $xy$-plane. In this simple case, the Navier-Stokes equations admit the exact solution of the velocity profile given by~\cite{2003Sigalotti}
\begin{equation}\label{eq:V_t}
  u(z,t) = \frac{{F{d^2}}}{{8\upsilon }}\left( {1 - {{\left( {\frac{{2z}}{d}} \right)}^2}} \right) - \sum\limits_{n = 0}^\infty  {\frac{{4{{\left( { - 1} \right)}^n}F{d^2}}}{{\upsilon {\pi ^3}{{\left( {2n + 1} \right)}^3}}}}  \cdot \cos \left[ {\frac{{\left( {2n + 1} \right)\pi z}}{d}} \right] \cdot \exp \left[ { - \frac{{{{\left( {2n + 1} \right)}^2}{\pi ^2}\upsilon t}}{{{d^2}}}} \right] \ ,
\end{equation}
where $d$ is the separation of the plates, $\upsilon$ the kinematic viscosity and $F$ a driving force per unit mass. The parameter set for the Poiseuille flow is $\rho=8.0$, $k_BT=1.0$, $a=75.0k_BT/\rho$, $\gamma=4.5$, $\sigma=3.0$ and $r_c=r_{cw}=1.0$. The kinematic viscosity of the DPD fluid can be computed by running a periodic Poiseuille flow~\cite{2005Backer}, which gives $\upsilon=0.275$.

\begin{figure}[t!]
  \centering
  \includegraphics[width=0.95\textwidth]{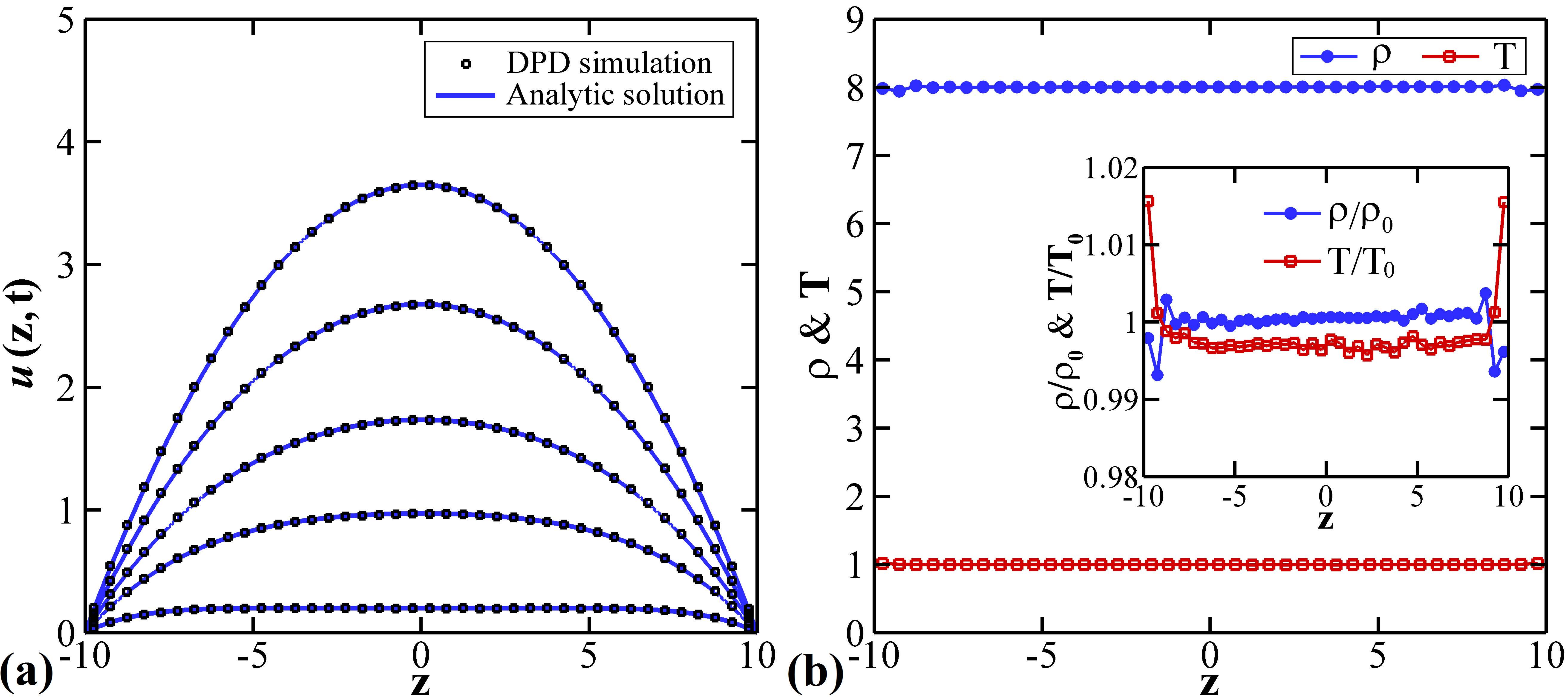}
  \caption{(a) Time evolution of the velocity profile $u(z,t)$ and comparison with analytic solution at $t=10,50,100,200$ and at steady state, and (b) density $\rho$ and temperature $T$ profiles in a Poiseuille flow. The inset of (b) shows negligible fluctuations of density ($<1\%$) and temperature ($<2\%$) in the vicinity of the wall surface. The simulation uses 24000 fluid particles and 4800 frozen particles for two flat walls of thickness $2.0$ in a computational domain of $30.0\times5.0\times24.0$ with $\rho=8.0, a=75.0k_BT/\rho, \sigma=3.0, k_BT=1.0$ and $r_c = r_{cw}=1.0$.}
  \label{fig:Poiseuille}
\end{figure}

More specifically, the DPD simulation of transient Poiseuille flow is performed in a computational domain of $30.0\times5.0\times24.0$ in DPD units, which contains 24000 fluid particles and 4800 frozen particles for solid walls with a thickness of $2.0$. The system is initialized with stationary fluid and two stationary walls. Periodic boundary condition is applied in $x$- and $y$-directions and no-slip boundary condition in $z$-direction. Then, a body force $g_x=0.02$ is applied on each DPD particle to drive the fluid, which is equivalent to imposing a pressure drop of $\rho g_x L_x$ on the channel of length $L_x$. To extract the velocity profile from the DPD simulation, we divide the computational domain into 48 bins of width $\Delta=0.5$ along the $z$-direction. The transient velocity profiles at $t=10,50,100,200$ and at steady state are plotted in Fig.~\ref{fig:Poiseuille}(a), where all local flow properties including particle density and kinetic temperature are obtained by averaging enough sampled data from 100 independent simulations initialized with different random seeds. The first and last bins contain both fluid and solid volumes because of the roughness of the wall surface, as shown in Fig.~\ref{fig:wall}. Considering the flat solid walls are made of randomly distributed particles, the volume of the raised part equals to the volume of the sunk part on average. Therefore, when we compute the density profile, all the fluid particles of $z<0.5$ are collected into the first bin and the fluid particles of $z>19.5$ are collected into the last bin. In Fig.~\ref{fig:Poiseuille}(a) we observe that the transient velocity profiles are in an excellent agreement with the analytical solution given by Eq.~\eqref{eq:V_t}, which indicates that the boundary method can provide accurate no-slip boundary condition on the wall surface. Furthermore, Fig.~\ref{fig:Poiseuille}(b) shows that our boundary method gives negligible density fluctuation (less than 1\%) and temperature fluctuation (less than 2\%) in the vicinity of the wall boundary.

\begin{figure}[t!]
  \centering
  \includegraphics[width=0.95\textwidth]{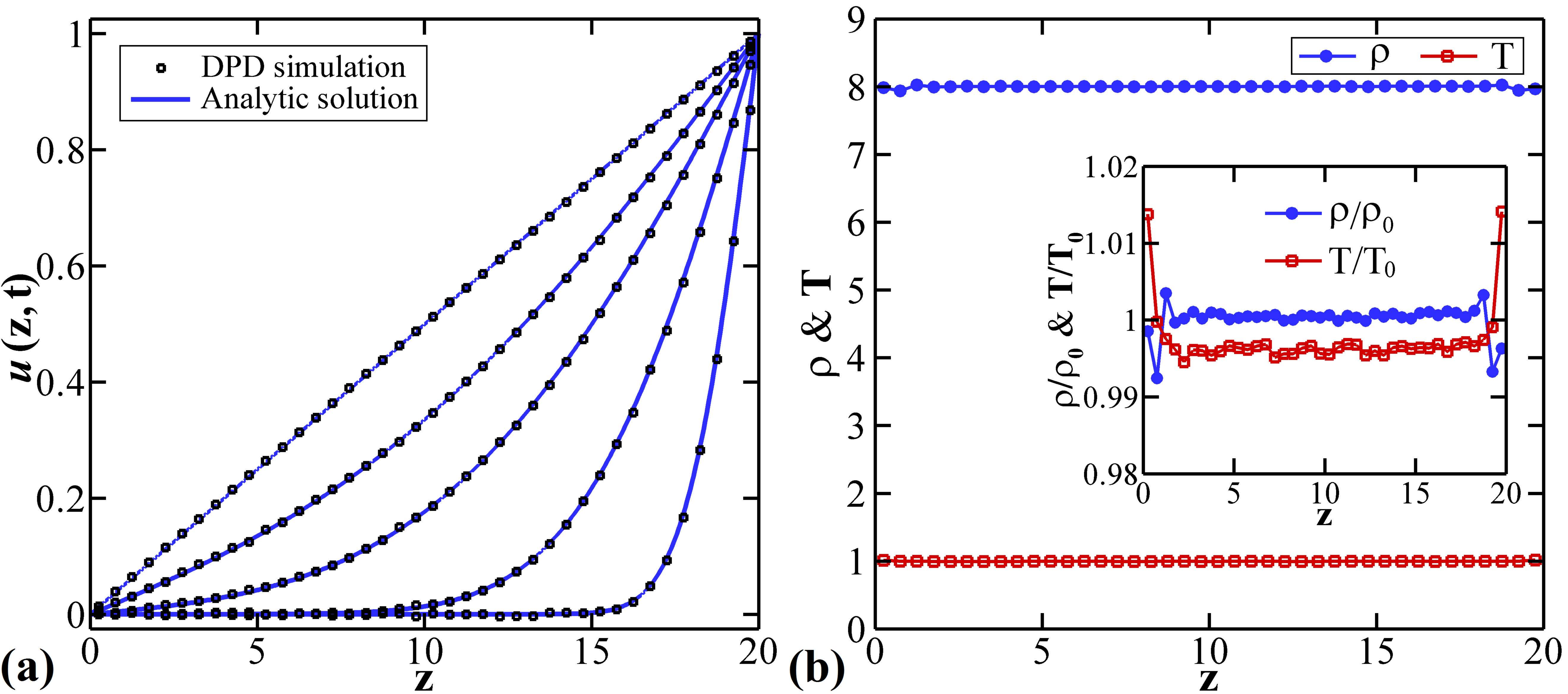}
  \caption{(a) Velocity $u$, (b) density $\rho$ and temperature $T$ profiles and comparison with Navier-Stokes solution in a Couette flow. The inset of (b) shows that the fluctuations of density and temperature in the vicinity of the wall surface are less than 2\%. The simulation uses 24000 fluid particles and 4800 solid particles for two flat walls with a thickness of $2.0$ in a computational domain of $30.0\times5.0\times24.0$. DPD parameters are set by $\rho=8.0, a=75.0k_BT/\rho, \sigma=3.0, k_BT=1.0$ and $r_c=r_{cw}=1.0$.}
  \label{fig:Couette}
\end{figure}

The next test case is used to validate the boundary method for moving flat walls. The Couette flow considers a viscous DPD fluid between two parallel plates, one of which is moving relative to the other. To simplify the case, we make the upper wall moving and the lower wall stationary. Similarly to the first test case, the DPD simulation of the Couette flow is performed in a computational domain of
$30.0\times5.0\times24.0$ with periodic boundary conditions in $x$- and $y$-directions and no-slip solid walls in $z$-direction. By solving a one-dimensional Navier-Stokes equation with boundary conditions of $u(0,t)=0$ and $u(20,t)=1.0$, an analytical solution for the transient velocity profile $u(z,t)$ can be obtained~\cite{2015ZLi_tDPD}
\begin{equation}\label{eq:Couette}
  u(z,t) = \frac{z}{d}U_0 + {\frac{2U_0}{\pi} \sum\limits_{n = 1}^\infty \frac{(-1)^n}{n} \sin\left(\frac{n\pi}{d} z \right) \exp\left(- \frac{n^2 \pi^2}{d^2}\nu t\right) } \ ,
\end{equation}
where $U_0$ is the velocity of the moving wall, $d$ is the separation between two plates and $\nu$ the kinematic viscosity of the fluid. The computational domain is divided into 48 bins of width $\Delta=0.5$ along the $z$-direction for obtaining local velocity profiles and fluid properties. Figure~\ref{fig:Couette}(a) shows a comparison between the transient velocity profiles obtained by DPD simulation and the theoretical solution of Eq.~\eqref{eq:Couette} at several times and also the steady state solution. The results are in good agreement, which validates the proposed boundary method for imposing the correct no-slip boundary condition for moving walls in the DPD simulation. Furthermore, similarly to the test case of Poiseuille flow, in Fig.~\ref{fig:Couette}(b) we observe negligible density fluctuation (less than 1\%) and temperature fluctuation (less than 2\%) in the vicinity of wall boundary.

\begin{figure}[t!]
  \centering
  \includegraphics[width=0.9\textwidth]{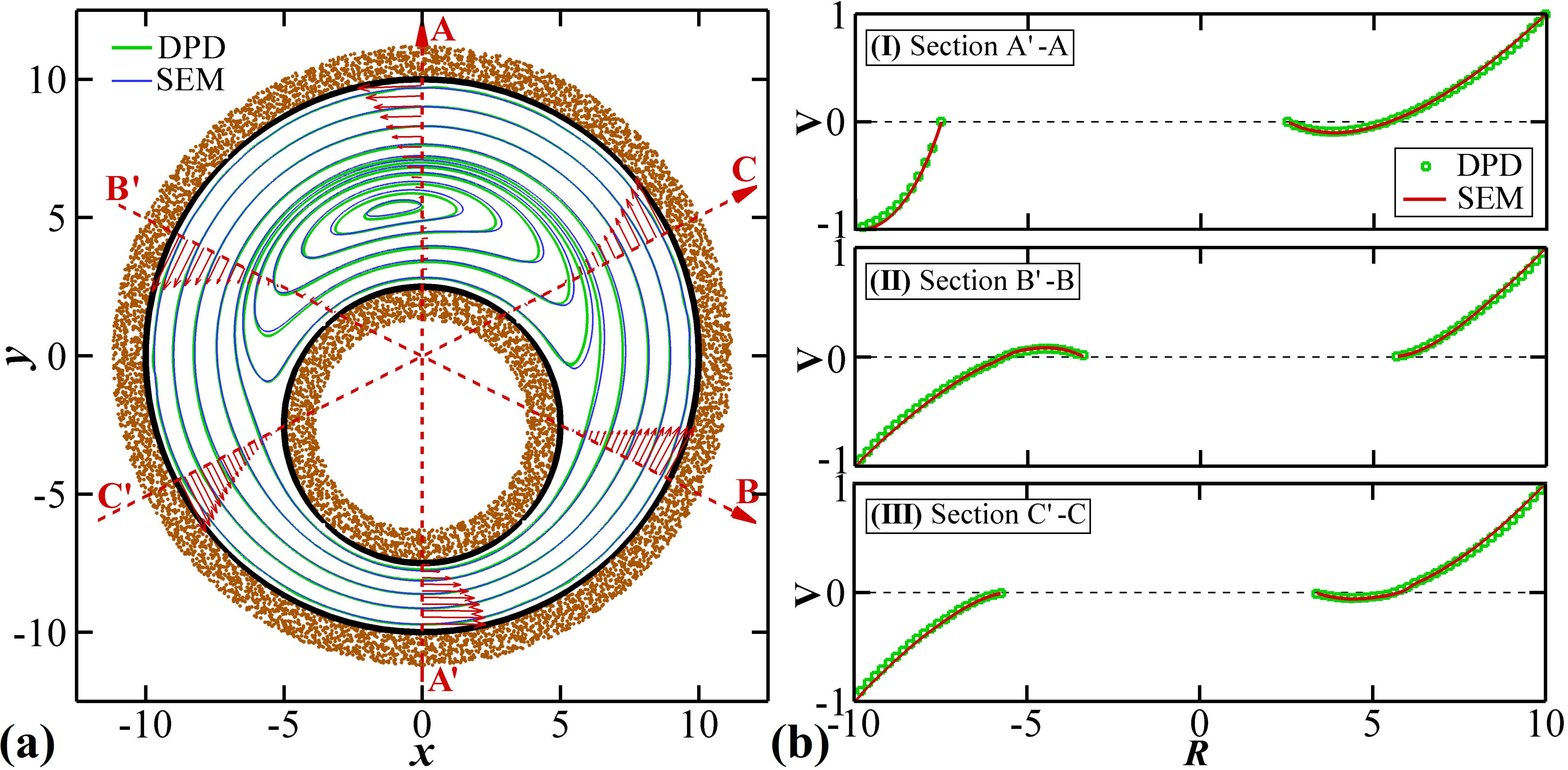}
  \caption{(a)Streamlines of the Wannier flow and (b) velocity profiles of sections A'-A, B'-B and C'-C obtained by the DPD simulation and the spectral element method (SEM). The DPD simulation uses 18850 fluid particles and 9054 solid particles for two cylinders of thickness 1.2 in a computational domain of $22.4\times22.4\times10.0$. The wall surface is represented by the isosurface of $\phi=0.5$ in the DPD simulation.}
  \label{fig:Wannier}
\end{figure}

In the previous two cases, we have tested the performance of the arbitrary boundary method for both stationary and moving flat walls. It yields correct no-slip boundary and successfully prevents the liquid particles from penetrating into solid boundaries. The next test case is for curved wall boundaries. We consider the so-called Wannier flow~\cite{1950Wannier} of two eccentric rotating cylinders shown in Fig. \ref{fig:Wannier} for validating the boundary method on curved walls since it involves both concave and convex wall boundaries. For this problem we setup the system with cylinder radii of $R_{outer} = 10$ at center $C_{outer}=\{0,0,0\}$ and $R_{inner} = 5.0$ at center $C_{inner}=\{0,-2.5,0\}$. The outer cylinder is set to rotate with an angular velocity of $\Omega=R^{-1}_{outer}=0.1$ while the inner cylinder is stationary. Then, the velocity and acceleration on the outer cylinder surface are $\mathbf{U}=\{u,v,w\}=\{-y\Omega,x\Omega,0\}$ and $\mathbf{a}=\{a_x,a_y,a_z\}=\{\Omega^2x,\Omega^2y,0\}$, respectively, which will be used in the predictor-corrector algorithm given by Eq.~\eqref{eq:Estimate}.

To construct a DPD system for the Wannier flow, we first fill a computational domain of $22.4\times22.4\times10.0$ with 40141 randomly distributed DPD particles. The DPD parameters are set as $\rho=8.0, a=75.0k_BT/\rho, \sigma=3.0, k_BT=1.0$ and $r_c = r_{cw}=1.0$. Then, we relax the system by running a short DPD simulation for 1000 time steps to eliminate the initial configurational energy. Subsequently, the DPD particles with $10.0\le R=\sqrt{x^2+y^2}\le 11.2$ are defined as outer cylinder, and the DPD particles with $3.8\le R=\sqrt{x^2+y^2}\le 5.0$ are defined as inner cylinder, while particles with $5 < R=\sqrt{x^2+y^2}< 10.0$ are fluid particles. All other particles are removed from the system. Finally, the DPD system has 18850 fluid particles and two cylinders of thickness $1.2$ consisting of 9048 solid particles. We run the DPD simulation long enough to obtain a fully developed Wannier flow. In the simulation, the boundary surface is defined by the isosurface of $\phi=0.5$. We compare the streamlines of the Wannier flow in Fig.~\ref{fig:Wannier}(a) and the velocity profiles of sections A'-A, B'-B and C'-C in Fig.~\ref{fig:Wannier}(b) obtained by the DPD simulation with the result obtained by the spectral element method (SEM). Results show that the DPD simulation is in very good agreement with the solution of SEM and we do not observe wall penetration in the DPD simulation, which indicates that the proposed boundary method can be safely applied to problems involving curved boundaries.

\begin{figure}[t!]
  \centering
  \includegraphics[width=1.0\textwidth]{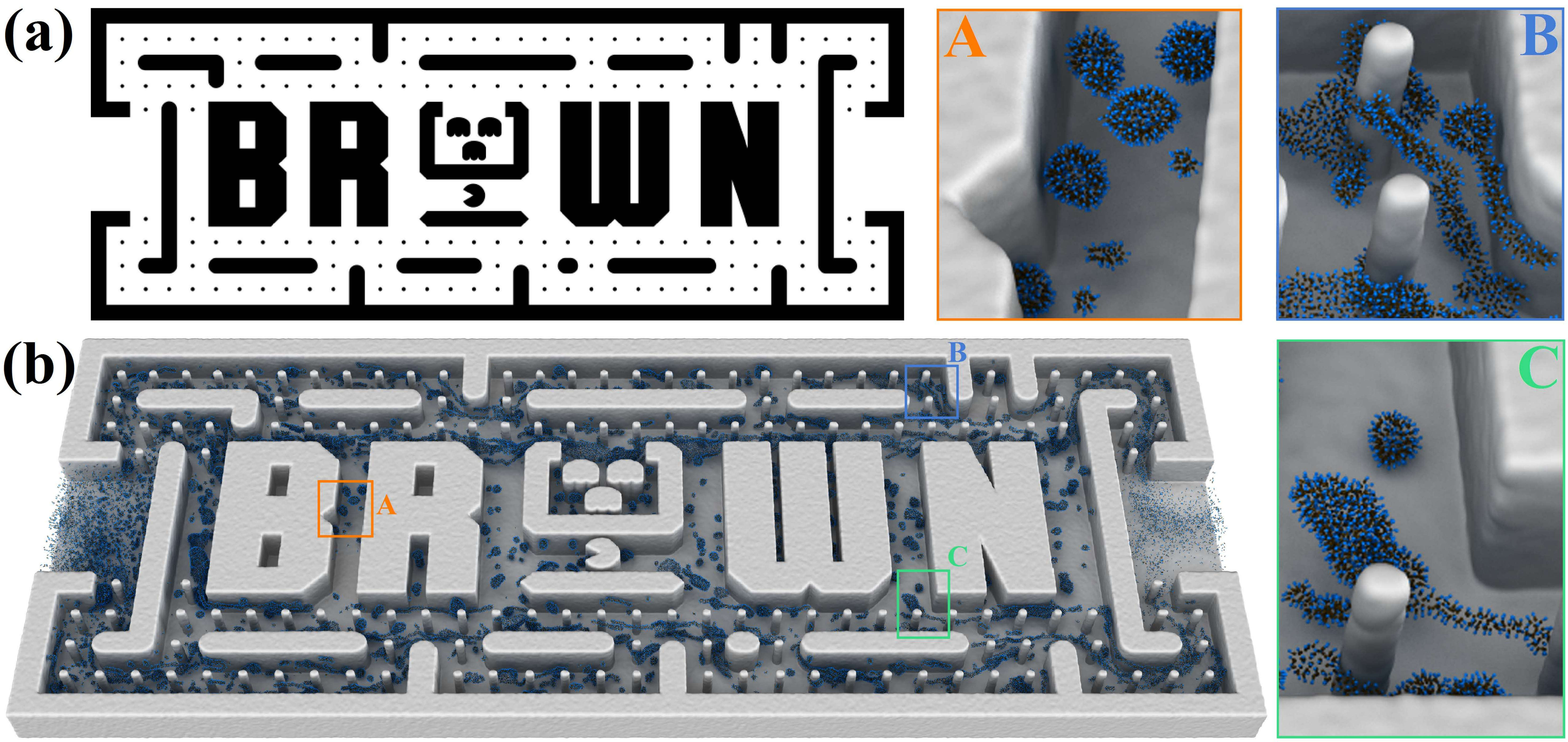}
  \caption{(a) The vector graphical image used for generating the DPD system of a ``Brown Pacman" microfluidic device. (b) Visualization of the surfactant solution flowing through the ``Brown Pacman" channel (see also Supporting Information for the movie). A, B and C are three zoom-in views of (b).}
  \label{fig:Pacman}
\end{figure}

To further demonstrate the capability of the presented arbitrary boundary method in realistic application scenarios, we construct a ``Brown Pacman" microfluidic device and carry out a simulation of a surfactant solution flowing through the microfluidic channel with complex geometry~\cite{2014M_Li}. The system is set up by mapping a vector graphics image of the desired channel geometry, as shown in Fig.~\ref{fig:Pacman}(a), onto a simulation box of size $600 \times 230 \times 24$ reduced units. DPD particles representing the channel wall are then placed randomly within regions with brightness $< 50\%$, while $6\,341\,124$ solvent particles and $300\,000$ surfactant particles with a volume concentration of 4.52\% are randomly placed in regions with brightness $> 50\%$. The system comprises of a total of $13\,248\,000$ DPD particles and the simulation is performed using the \usermeso GPU-accelerated DPD package \cite{2014Tang}. Each surfactant molecule has one hydrophilic bead (H) and one hydrophobic bead (T) connected by a harmonic bond with potential $E_b(r) = K ( r - r_0 )^2 $, where $K$ is the spring force constant, and $r$, $r_0$ the instantaneous and equilibrium bond length. A cutoff distance $r_c=1.0$ is used for the pairwise interaction and $r_{cw}=1.0$ for the local detection method. The wall surface is represented by the isosurface of $\phi=0.5$. The interaction matrix between the surfactant, solvent and wall particles is given in Table~\ref{table:aij}. A lateral pressure gradient, $-\partial p/\partial x = c ( v_x - v_x^0 )$, where $c = 0.25$ and $v_x^0 = 4$, is applied at the inlet of the channel to drive the flow. The system is first optimized using a short run of DPD simulation. A time step size of $\Delta t = 0.01$ is then used to simulate the system for $1 \times 10^6$ time steps.

In Fig.~\ref{fig:Pacman}(b), we observe rich phenomena of surfactant dynamically assembling and disassembling following the flow (see also Supporting Information for the movie). Three local zoom-in views of Fig.~\ref{fig:Pacman}(b) are shown in Fig.~\ref{fig:Pacman}A, B and C. More specifically, zone A is located between the walls of ``B" and ``R", where the flow field is almost stationary. Consequently, the surfactant molecules in a shear-free solution self-assemble into small spherical aggregates, as shown in Fig.~\ref{fig:Pacman}A. However, Fig.~\ref{fig:Pacman}B shows that the surfactant molecules form elongated wormlike micelles under strong shear flow. We observe that these wormlike micelles flow around the small cylinders without wall penetration. Zone C is located at a transition area from a nearly stationary flow to a shear flow, where a shear-induced phase transition from spherical micelles to elongated wormlike micelles is shown in Fig.~\ref{fig:Pacman}C. Since the boundary geometry is computed on-the-fly, the proposed local detection method takes care of imposing no-slip boundary conditions and preventing wall penetrations automatically, even for such a complex microfluidic device. This may be very valuable for many realistic applications.

\begin{table}[htb!]
\centering
\caption{Repulsive force constants $a_{ij}$ for microfluidic channel}
\label{table:aij}
\begin{tabular}{p{1.5cm}p{1cm}p{1cm}p{1.5cm}p{1cm}}
\hline\hline
        & H                      & T    & Solvent                 & Wall                    \\
\hline
H       & 45                     & 75   & 37.5                    & 150                     \\
T       & 75                     & 37.5 & 150                     & 150                     \\
Solvent & 37.5                   & 150  & 37.5                    & 37.5                    \\
Wall    & 150                    & 150  & 37.5                    & 37.5                    \\
\hline\hline
\end{tabular}
\end{table}

\section{Summary and Discussions}\label{sec:4}
A local detection method tackling the challenges induced by arbitrarily shaped boundaries and complex geometries in dissipative particle dynamics (DPD) simulations has been proposed. By computing a boundary volume fraction (BVF) for each fluid particle, the solid boundary is detected on-the-fly by the fluid particles according to local particle configuration. At a small extra computational cost, the fluid particles become autonomous to find the wall surface and to infer the wall penetration based on the coordinates of their neighboring particles. A predictor-corrector algorithm was employed to prevent the fluid particles from penetrating into the wall boundaries, and the effective dissipative coefficients for liquid-solid interactions were used to impose no-slip boundary condition on the wall surface.

We employed randomly distributed particles to represent walls to allow easiness and generality for construction of DPD systems involving arbitrary-shape boundaries. Theoretically, to the limit of the continuum approximation, the wall surface can be accurately represented by the isosurface of BVF $\phi=0.5$ when the solid particles are dense enough. However, in practical implementations, the random distribution of discrete particles with finite number density will introduce surface roughness of wall boundaries, which comes from the estimation error of BVF based on the discrete particles. We demonstrated that the magnitude of roughness associated with the randomness of particle distribution is monotonically controllable by tuning the cutoff radius for computing BVF and the number density of DPD particles. Since any natural solid wall contains more or less chemical/physical heterogeneities on the surface, especially at the mesoscopic scale, such heterogeneity can be modeled qualitatively by the roughness on walls in the DPD systems. In this respect, the proposed boundary method provides a convenient way to generate various sizes of the roughness for representation of different degrees of the heterogeneity and to introduce curvature-dependent slip for hydrodynamics as discussed in the~\ref{sec:app}.

The transition Poiseuille and Couette flows as well as the Wannier flow were used as benchmark tests for verifying the proposed arbitrary boundary method. The results showed that the proposed boundary method imposes the correct no-slip boundary condition for both stationary and moving walls in the DPD simulation, and yields negligible density fluctuation (less than 1\%) and temperature fluctuation (less than 2\%) in the vicinity of wall surface. To further demonstrate the capability of the presented arbitrary boundary method in realistic application scenarios, a ``Brown Pacman" microfluidic device with complex geometry was constructed directly from a vector graphics image and a DPD simulation of surfactant solution flowing through this complex microfluidic device was carried out. The validity of this boundary method is confirmed by examining the rich dynamics of surfactant micelles flowing around the small cylinders without wall penetration.

Since this local detection method only uses local information of neighboring particles for computing the value of BVF and satisfies designed boundary conditions on-the-fly, it provides a practical and efficient way to deal with complex geometries and impose the no-slip boundary condition on wall surface in DPD simulations. With the local detection method, it is no longer necessary to mathematically define the boundary geometry for DPD simulations, which enables us to construct DPD systems directly from experimental CT images or computer-aided designs/drawings. Moreover, this method is not only valuable for stationary arbitrary-shape boundaries, but also for the moving boundaries and deformable boundaries.

Although we presented here that the surface roughness is controllable by varying the number of neighboring particles, this boundary method cannot accurately capture large curvatures of wall boundary where the radius of curvature is too small to be identified from the surface roughness. To this end, higher resolution of DPD system is required to represent the large curvature properly so that the size of random surface roughness is much smaller than the radius of curvature.

\section*{Acknowledgements}
This work was primarily supported by the DOE Center on Mathematics for Mesoscopic Modeling of Materials (CM4). This work was also sponsored by the U.S. Army Research Laboratory and was accomplished under Cooperative Agreement Number W911NF-12-2-0023 to University of Utah. An award of computer time was provided by the Innovative and Novel Computational Impact on Theory and Experiment (INCITE) program with the resources of the Argonne Leadership Computing Facility (ALCF) and the resources of the Oak Ridge Leadership Computing Facility (OLCF). Z. Li would like to thank Dr. Yue Yu for her support on running the spectral element simulation of the Wannier flow.


\appendix
\section{Error Analysis for Curved Surfaces}\label{sec:app}
\begin{figure}[b!]
  \centering
  \includegraphics[width=0.8\textwidth]{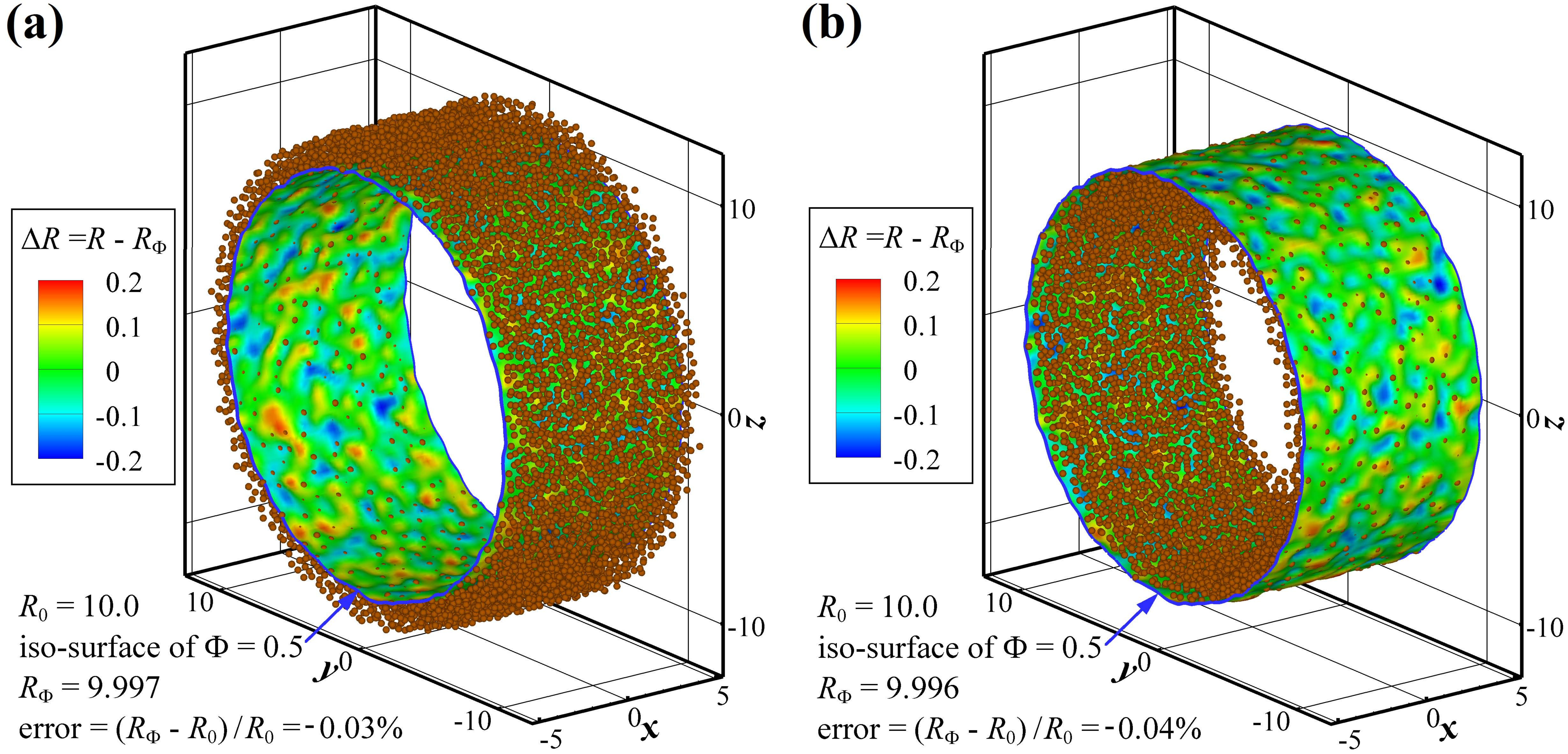}
  \caption{Performance accuracy of representing curved wall boundaries, including (a) concave surface and (b) convex surface, by the isosurface of $\phi=0.5$. The expected radii of cylinders are $R_0=10.0$, the isosurface of $\phi=0.5$ gives $R_{\Phi}=9.997$ for the concave cylinder and $R_{\Phi}=9.996$ for the convex cylinder. Spherules represent randomly distributed solid particles constituting cylinders.}
  \label{fig:cylinder}       
\end{figure}

Representing the wall boundary by the isosurface of $\phi=0.5$ is accurate for flat wall boundaries, as indicated by Eq.~\eqref{eq:phi_h}. However, for curved boundaries, the isosurface of $\phi=0.5$ will deviate from the designed boundary surface. To analyze the error and the performance of representing curved wall boundaries with the isosurface of $\phi=0.5$, we construct a series of concave and convex cylinders to quantify the error induced by the curvature of wall boundaries. Fig.~\ref{fig:cylinder} shows the geometry of concave and convex cylinders. The expected radii of both cylinders are $R_0=10.0$. Let $\rho_w=8.0$ be the particle number density, the average separation between particles is $\delta=\rho_w^{-1/3}=0.5$. We fill a computational domain of $10.0\times25.0\times25.0$ with 50000 randomly distributed DPD particles. Then, we relax the system by running a short DPD simulation to eliminate initial randomicity. For the concave cylinder, the DPD particles with $R=\sqrt{y^2+z^2}$ ranging from $10.0$ to $12.0$ are defined as solid particles, as shown in Fig.~\ref{fig:cylinder}(a). Similarly, the convex cylinder is made up by the discrete DPD particle with $R=\sqrt{y^2+z^2}$ ranging from $8.0$ to $10.0$.

\begin{figure}[h!]
  \centering
  \includegraphics[width=0.8\textwidth]{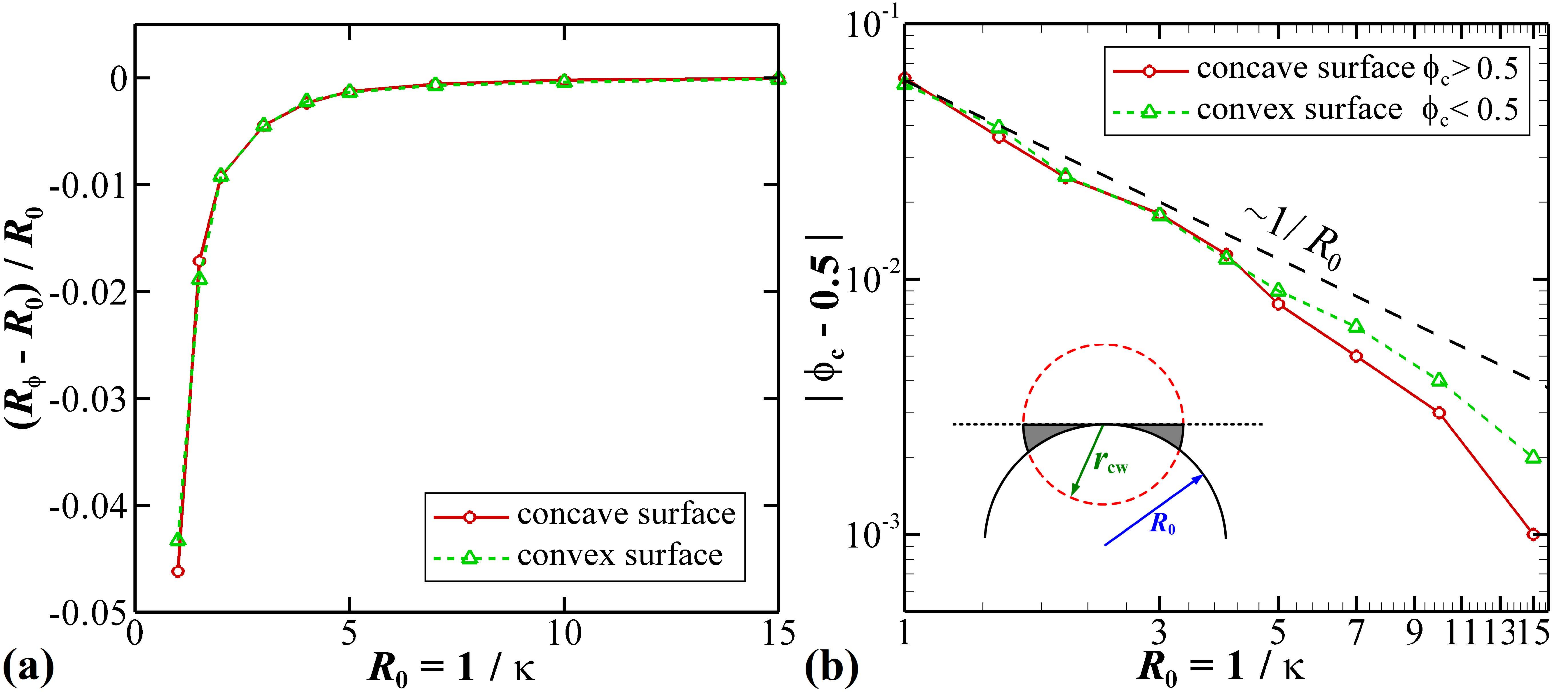}
  \caption{(a) Relative error $(R_{\Phi}-R_0)/R_0$ vs radius of curvature of the curved wall boundaries $R_0=1/\kappa$ in which $\kappa$ is the curvature. (b) Corrected magnitude of boundary friction function $\phi_c$ that accurately represents the expected concave and convex wall boundaries.}
  \label{fig:error}       
\end{figure}

By defining $r_{cw}=1.0$, the value of $\varphi(\mathbf{r})$ can be easily computed using Eq.~\eqref{eq:Phi}. Figure~\ref{fig:cylinder} shows the performance of representing concave and convex boundaries by the isosurfaces of $\phi=0.5$. The designed radii of both two cylinders are $R_0=10.0$, while the iso-surface of $\phi=0.5$ gives $R_{\Phi}=9.997$ for the concave cylinder and $R_{\Phi}=9.996$ for the convex cylinder. Although the isosurfaces of $\phi=0.5$ slightly deviate from the expected cylinder radius, the relative error $(R_{\Phi}-R_0)/R_0$ are negligible for $R_0=10.0$, i.e., less than 0.05\%.

Let $\kappa$ be the curvature of the cylinder surface, as the radius of curvature $R_0=1/\kappa$ decreases, the relative error $(R_{\Phi}-R_0)/R_0$ increases, as plotted in Fig.~\ref{fig:error}(a). We observe that the relative error $(R_{\Phi}-R_0)/R_0$ is about $4.5\%$ for $R_0=1.0$, however, it becomes less than $1\%$ for $R_0\ge 2.0$. Consequently, using the isosurface of $\phi=0.5$ to represent the curved boundaries does not induce significant error for small curvatures, i.e., $\kappa=R_0^{-1} \le 0.5$.

Theoretically, a corrected magnitude of the boundary friction function $\phi_c$ rather than $0.5$ can be defined to accurately represent the curved boundaries. Using the continuum approximation, the extra volume for curved surfaces different from a flat surface is $\pi r_{cw}^4/4R_0$, as shown by the dark domain in the inset of Fig.~\ref{fig:error}(b). Then, the corrected boundary friction functions are
$\phi_c = 0.5 + 3r_{cw}/16R_0$ for concave surfaces and $\phi_c = 0.5 - 3r_{cw}/16R_0$ for convex surfaces. Therefore, the difference between $\phi_c$ and $0.5$ decreases as the radius of curvature $R_0=1/\kappa$ increases, which is given by
\begin{equation}\label{eq:Phi_c}
  \left|\phi_c - 0.5 \right| = \frac{\pi r_{cw}^4/4R_0}{4\pi r_{cw}^3/3} = \frac{3r_{cw}}{16R_0}\ .
\end{equation}
The formula of Eq.~\eqref{eq:Phi_c} is obtained based on the continuum approximation. For the wall consisting of discrete DPD particles, we observe in Fig.~\ref{fig:error}(b) that the decay of $\left|\phi_c - 0.5 \right|$ is slightly faster than $R_0^{-1}$. This is because the distribution of DPD particle is initially regularized by performing a short simulation.

It is worth noting that employing the isosurface of $\phi=0.5$ to represent the curved boundaries could introduce curvature-dependent slip for hydrodynamics, which would be practically useful for some applications where partial slips on curved wall surfaces are expected. As described in Fig.~\ref{fig:error}(a), the isosurface of $\phi=0.5$ deviates from the real curved surface and the deviation increases as the curvature increases. Since the arbitrary boundary method imposes no-slip boundary condition on the isosurface of $\phi=0.5$, for both concave and convex geometries the real boundary surface will have partial slip for velocity, where the slip length depends on the curvatures of the boundary as shown in Fig.~\ref{fig:error}(a).

\end{document}